\documentclass[letterpaper, english, preprint, aps, prd, nofootinbib, showpacs]{revtex4-1}
\usepackage[T1]{fontenc}
\usepackage[latin9]{inputenc}
\usepackage{babel}
\usepackage{verbatim}
\usepackage{textcomp}
\usepackage{amsmath}
\usepackage{amssymb}
\usepackage{graphicx}
\usepackage{makecell}
\usepackage{subcaption}
\usepackage{color}
\usepackage[unicode=true,pdfusetitle,
 bookmarks=true,bookmarksnumbered=false,bookmarksopen=false,
 breaklinks=false,pdfborder={0 0 1},backref=false,colorlinks=false]
 {hyperref}
\makeatletter

\pdfpageheight\paperheight
\pdfpagewidth\paperwidth

\usepackage{chngcntr}
\counterwithout{figure}{section}

\makeatother

\begin{document}

\title{Towards the nucleon hadronic tensor from lattice QCD}

\author{\texorpdfstring{Jian Liang$^{1,}$\footnote{jian.liang@uky.edu}, Terrence Draper$^{1}$, Keh-Fei Liu$^{1,}$\footnote{liu@g.uky.edu}, Alexander Rothkopf$^{2}$ and Yi-Bo Yang$^{3}$}}

\affiliation{$^{1}$\mbox{Department of Physics and Astronomy, University of Kentucky, Lexington, KY 40506, USA} 
$^{2}$\mbox{Faculty of Science and Technology, University of Stavanger, 4021 Stanvanger, Norway}
$^{3}$\mbox{Institute of Theoretical Physics, Chinese Academy of Sciences, Beijing 100190, China}
\\~\\
\includegraphics[scale=0.12]{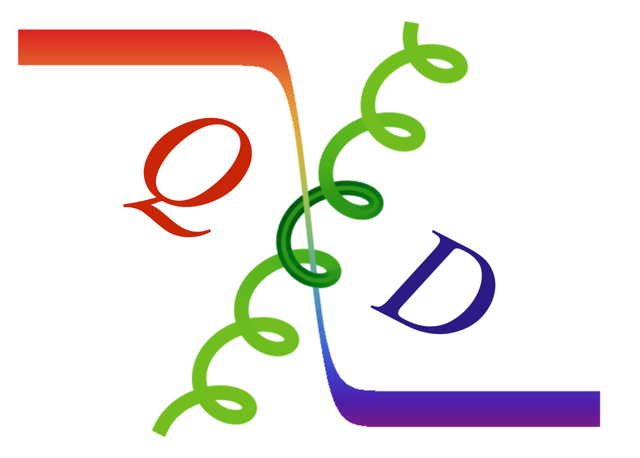}}

\collaboration{$\chi$QCD Collaboration}\noaffiliation

\begin{abstract}
We present the first calculation of the hadronic tensor on the lattice for the nucleon.
The hadronic tensor can be used to extract the structure functions in deep inelastic scatterings and also 
provide information for the neutrino-nucleon scattering which is crucial to the neutrino-nucleus scattering experiments at low energies.
The most challenging part in the calculation is to solve an inverse problem. We have implemented and tested three algorithms using mock data, showing that the Bayesian Reconstruction method has the
best resolution in extracting peak structures while the Backus-Gilbert and Maximum Entropy methods are somewhat more stable for the flat spectral function.
Numerical results are presented for both the elastic case (clover fermions on domain wall configuration with $m_\pi\sim$ 370 MeV and $a\sim$ 0.06 fm) and a case (anisotropic clover lattice 
with $m_\pi\sim$ 380 MeV and $a_t\sim$ 0.035 fm) with large momentum transfer.
For the former case, the reconstructed Minkowski hadronic tensor gives precisely the vector charge which proves the feasibility of the approach.
While for the latter case, the nucleon resonances and possibly shallow inelastic scattering contributions around $\nu=1$ GeV are clearly observed but no information is obtained 
for higher excited states with $\nu>2$ GeV.
A check of the effective masses of $\rho$ meson with different lattice setups indicates that, in order to reach higher energy transfers,
using lattices with smaller lattice spacings is essential. 

\end{abstract}
\maketitle

\section{\label{sec:Introduction}Introduction}

In scattering processes involving
nucleons such as deep inelastic scattering
(DIS) and neutrino-nucleon scattering at low energies, the hadronic tensor $W_{\mu\nu}$ is used to characterize the nonperturbative
nature of the nucleon structure. It is the imaginary part
of the forward virtual Compton scattering amplitude $W_{\mu\nu}=\frac{1}{2\pi}{\rm Im}T_{\text{\ensuremath{\mu\nu}}}$
and can be expressed as a nucleon matrix element with two current
operators inserted
\begin{equation}
\label{HT1}
W_{\mu\nu}=\frac{1}{4\pi}\int d^{4}ze^{iq\cdot z}\left\langle p,s\left|\left[J_{\mu}^{\dagger}(z)J_{\nu}(0)\right]\right|p,s\right\rangle .
\end{equation}
The hadronic tensor of the vector currents can be further decomposed, according to its Lorentz
structure, into structure functions, i.e.,
\begin{equation}
{W_{\mu\nu}}=\left(-g_{\mu\nu}+\frac{q_{\mu}q_{\nu}}{q^{2}}\right){F_{1}(x,Q^{2})}+\frac{\hat{p}_{\mu}\hat{p}_{\nu}}{p\cdot q}{F_{2}(x,Q^{2})}
\end{equation}
for the unpolarized case, where $\hat{p}_{\mu}=p_{\mu}-\frac{p\cdot q}{q^{2}}q_{\mu}$. $p_\mu$ and $q_\mu$ are
the nucleon 4-momentum and momentum transfer, respectively. $x$ is the Bjorken $x=\frac{Q^2}{2p\cdot q}$.
The structure functions are valuable quantities which reveal the
inner structure of the nucleon. They can be used to extract parton
distribution functions (PDF's) through the QCD factorization theorem
$F_{i}=\sum_{a}c_{i}^{a}\otimes f_{a}$, where the convolution kernel
$c_{i}^{a}$ is perturbatively calculable.

Due to their nonperturbative nature and importance, it is natural
to explore the possibility of calculating the hadronic tensor and structure functions
with Lattice QCD, which is a first-principle nonperturbative
method of solving the strong interaction. 
While the proposal for a lattice QCD based evaluation
has been put forward more than 20 years ago~\citep{Liu:1993cv,Liu:1999ak}, it has only recently
become feasible to compute the necessary 4-point correlation functions,
thanks to the increases in computing power~\cite{Liu:2016djw,Liang:2017mye}.
In recent years, there has been a lot
of effort in the lattice community focusing on the computation of
$x$-dependent PDF's. Examples are Quasi-PDFs and LaMET \citep{Ji:2013dva,Lin:2018qky},
Compton amplitude \citep{Chambers:2017dov}, Pseudo-PDFs \citep{Radyushkin:2017cyf,Orginos:2017kos} and
Lattice cross sections \citep{Ma:2017pxb,Sufian:2019bol}. Each of
the approaches has its own advantages and difficulties. Since the hadronic
tensor is scale-independent and the structure functions are frame-independent,
the lattice calculation of the hadronic tensor has its unique advantages in that 
no renormalization nor large nucleon momentum are needed.
However, to convert the hadronic tenor from Euclidean space to Minkowski space
involves an inverse problem~\cite{Liu:2016djw} which presents a substantial numerical challenge.
In case that reliable lattice results for the Minkowski hadronic tensor are obtained for a set of 
kinematic setups, they can be used to obtain parton distribution functions via the factorization theorem,
as are carried out in global fittings of experiments.

Another feature of calculating the hadronic tensor on the lattice
is that it reveals explicitly the connected-sea anti-parton contribution.
It has been pointed out~\citep{Liu:1993cv,Liu:2012ch} that the Gottfried
sum rule violation (i.e., the $\bar{u}$ and $\bar{d}$ difference of PDF's)
can be explained by the existence of the connected-sea anti-partons.
Recently, the ratio
of strange to $u/d$ momentum fraction in disconnected insertions was calculated~\citep{Liang:2019xdx},
which helps to separate the connected and disconnected-sea
parton distributions in global fittings. The calculation
of hadronic tensor using Euclidean 4-point correlation functions will provide
a direct proof of the degrees of freedom of connected-sea anti-partons
and can finally resolve the puzzle of Gottfried sum rule violation.

In addition to deep inelastic scattering, the hadronic
tensor plays an important role too for scatterings at lower energies.
One example is the experiments
of neutrino-nucleus scattering, e.g. LBNF/DUNE \citep{Acciarri:2015uup}
at Fermilab, which aims to study the neutrino properties. 
These experiments face several challenges like the reconstruction of the neutrino beam energy and flux and
the consideration of the nuclear effects and models~\cite{Alvarez_Ruso_2018}.
In view of this, the input of accurate determination of the neutrino-nucleon scattering is
vital to investigating the nuclear effects of neutrino-nucleus scattering.
However, it is not trivial to study the neutrino-nucleon scattering
since at different
beam energies, different contributions (elastic (EL), resonance
(RES), shallow inelastic (SIS) and DIS) dominate the total cross
section \citep{Formaggio:2013kya}.
Nevertheless, the hadronic tensor is useful
in all the energy regions. For example,
in the EL region of neutrino-nucleon scattering which is relevant to the quasi-elastic neutrino-nucleus scattering, the hadronic
tensor is actually the square of the elastic form factors of the nucleon and as a result, the
cross section of the neutrino-nucleus scattering can be calculated by
combining the nucleon form factors and nuclear models about the nucleon
distribution inside a nucleus. In the RES, SIS and DIS regions, inelastic
neutrino-nucleon scatterings emerge and one will need to have the
hadronic tensor to cover all the inclusive contributions.
In this sense, calculating the hadronic tensor is so far the only
way we know that Lattice QCD can serve the neutrino experiments in the whole energy range~\cite{Kronfeld:2019nfb}\footnote{We thank A. Kronfeld for bringing to our attention the relevance of the hadronic tensor
to the neutrino-nucleon scattering.}.

Since Lattice QCD is formulated with Euclidean time and the
hadronic tensor involves a 4-dimensional Fourier transform, one cannot
calculate the hadronic tensor directly on the lattice. Instead, we
calculate its counter part in Euclidean space and then convert
it back to Minkowski space. The formalism of Euclidean hadronic
tensor is discussed in Sec.~\ref{sec:form}. The conversion to the
Minkowski space is implemented by solving the inverse problem which
is the most challenging part of our calculation. We will discussed
three methods and two examples in Sec.~\ref{sec:Inverse}. 
Sec.~\ref{sec:results} presents numerical results for both the elastic case and a case with large momentum transfer.
Discussion on the results comes in Sec.~\ref{sec:discussion}.

\section{\label{sec:form}Lattice Formalism of hadronic tensor}

After inserting a complete set of intermediate states in Eq.~(\ref{HT1}) and carrying out
the integral, one comes to the following expression of the hadronic tensor~\cite{Liu:2016djw}:
\begin{equation}
\label{MHT}
W^M_{\mu\nu}=\frac{1}{2}\sum_{n}\int\prod_{i}^{n}\left[\frac{d^{3}\vec{p}_{i}}{(2\pi)^{3}2E_{i}}\right]\langle p,s|J_{\mu}^{\dagger}(0)|n\rangle\langle n|J_{\nu}(0)|p,s\rangle(2\pi)^{3}\delta^{4}(q-p_{n}+p),
\end{equation}
where $q$ is the momentum transfer, $p$ is the nucleon momentum
and $p_{n}$ is the momentum of the $n$'th intermediate state. The
4-dimensional Dirac delta function ensures the conservation of 4-momentum
and picks out the contribution of a particular momentum transfer.
However, it is noted in~\cite{Liu:2016djw} that if one carries out the integral in Eq.~(\ref{HT1}) in the Euclidean case,
the Fourier transform in the time direction becomes a Laplace transform
\begin{equation}
\label{Laplace}
W'_{\mu\nu}=\frac{1}{4\pi}\sum_{n}\int dte^{\left(\nu-(E_{n}-E_{p})\right)t}\int d^{3}\vec{z}e^{i\vec{q}\cdot\vec{z}}\langle p,s|J_{\mu}^{\dagger}(\vec{z})|n\rangle\langle n|J_{\nu}(0)|p,s\rangle,
\end{equation}
and after the integration one has
\begin{equation}
\label{Laplace2}
W'_{\mu\nu}=\frac{1}{4\pi}\sum_{n}\frac{e^{\left(\nu-(E_{n}-E_{p})\right)T}-1}{\nu-(E_{n}-E_{p})}\int d^{3}\vec{z}e^{i\vec{q}\cdot\vec{z}}\langle p,s|J_{\mu}^{\dagger}(\vec{z})|n\rangle\langle n|J_{\nu}(0)|p,s\rangle,
\end{equation}
where $\nu$ is the energy transfer, $E_{p}$ and $E_{n}$ are the
energies of the external nucleon and the $n$'th intermediate state, and
$T$ is the integration length in the time direction. The factor
of $\frac{e^{\left(\nu-(E_{n}-E_{p})\right)T}-1}{\nu-(E_{n}-E_{p})}$
is problematic since it does not converge if $\nu-(E_{n}-E_{p})>0$.
This happens when $\nu$ is greater that the energy gap between the nucleon and the intermediate state, such as $\Delta$, Roper, ... .
Besides, even if the numerator converges, $\frac{1}{\nu-(E_{n}-E_{p})}$
is not a good approximation to the delta function and one cannot pick
out the clean contribution for a specific $\nu$ as nearby states will mix.

Instead, we construct the following 4-point correlation function with
only a 3-dimensional Fourier transform
\begin{equation}
C_{4}(t_{f},t_{2},t_{1})=\sum_{\vec{x}_{f}}e^{-i\vec{p}\cdot\vec{x}_{f}}\sum_{\vec{x}_{1}\vec{x}_{2}}e^{-i\vec{q}\cdot(\vec{x}_{2}-\vec{x}_{1})}\left\langle \chi_{N}(\vec{x}_{f},t_{f})J_{\mu}^{\dagger}(\vec{x}_{2},t_{2})J_{\nu}(\vec{x}_{1},t_{1})\bar{\chi}_{N}(\vec{0},t_{0})\right\rangle ,
\end{equation}
and the normal nucleon 2-point function as
\begin{equation}
C_{2}(t_{f})=\sum_{\vec{x}_{f}}e^{-i\vec{p}\cdot\vec{x}_{f}}\left\langle \chi_{N}(\vec{x}_{f},t_{f})\bar{\chi}_{N}(\vec{0},t_{0})\right\rangle .
\end{equation}
Then, the Euclidean hadronic tensor ${W}^E_{\mu\nu}(\vec{p},\vec{q},\tau)$ is defined by the ratio of the 4-point
function to the 2-point function~\cite{Liu:1993cv,Liu:1999ak,Liu:2012ch,Liu:2016djw,Liang:2017mye}
\begin{equation}
\label{eq:EHT}
{W}^E_{\mu\nu}(\vec{p},\vec{q},\tau)=\frac{E_{p}}{m_{p}}\frac{{\rm Tr}[\Gamma_{e}C_{4}]}{{\rm Tr}[\Gamma_{e}C_{2}]}\to\sum_{\vec{x}_{1}\vec{x}_{2}}e^{-i\vec{q}\cdot(\vec{x}_{2}-\vec{x}_{1})}\langle p,s|J_{\mu}^{\dagger}(\vec{x}_{2},t_{2})J_{\nu}(\vec{x}_{1},t_{1})|p,s\rangle,
\end{equation}
where $E_{p}$ and $m_{p}$ are the energy and mass of the nucleon.
We can insert the intermediate states again between the two currents
and we have
\begin{equation}
{W}^E_{\mu\nu}=\sum_{n}A_{n}e^{-(E_{n}-E_{p})\tau},
\end{equation}
where 
\begin{equation}
A_{n}\equiv\sum_{\vec{x}_{1}\vec{x}_{2}}e^{-i\vec{q}\cdot(\vec{x}_{2}-\vec{x}_{1})}\langle p,s|J_{\mu}^{\dagger}(\vec{x}_{2},0)|n\rangle\langle n|J_{\nu}(\vec{x}_{1},0)|p,s\rangle
\end{equation}
and $\tau=t_{2}-t_{1}$.

\begin{figure}[!h]
\centering
\begin{subfigure}[b]{0.25\textwidth}
\includegraphics[width=\textwidth,page=1]{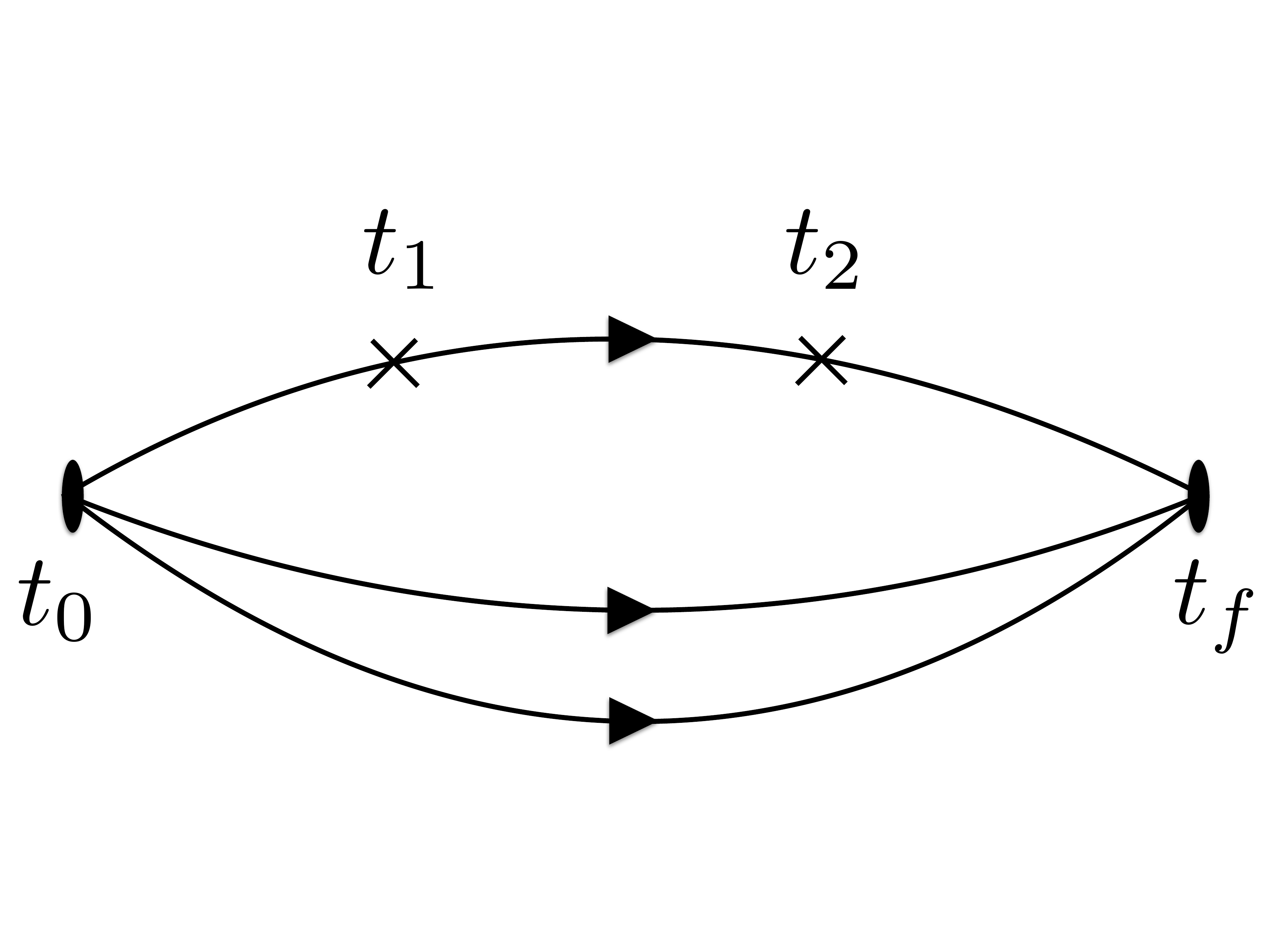}
\caption{\label{ca}}
\end{subfigure}
\begin{subfigure}[b]{0.25\textwidth}
\includegraphics[width=\textwidth,page=2]{draw}
\caption{\label{cb}}
\end{subfigure}
\begin{subfigure}[b]{0.25\textwidth}
\includegraphics[width=\textwidth,page=4]{draw}
\caption{\label{cc}}
\end{subfigure}

\begin{subfigure}[b]{0.25\textwidth}
\includegraphics[width=\textwidth,page=3]{draw}
\caption{\label{cd}}
\end{subfigure}
\begin{subfigure}[b]{0.25\textwidth}
\includegraphics[width=\textwidth,page=5]{draw}
\caption{\label{ce}}
\end{subfigure}
\begin{subfigure}[b]{0.25\textwidth}
\includegraphics[width=\textwidth,page=6]{draw}
\caption{\label{cf}}
\end{subfigure}

\caption{Topologically distinct diagrams in the Euclidean-path integral formulation of the nucleon hadronic tensor.
Figs.~\ref{ca}, \ref{cb} and \ref{cc} contain all twists and Figs.~\ref{cd}, \ref{ce} and \ref{cf} contain high twists only.
\label{contractions}}
\end{figure}

The computation of 4-point functions consumes the most computer resources in our calculation. 
Doing the Wick contraction for the 4-point function leads to 
several topologically distinct diagrams in the Euclidean path-integral formulation (shown in Fig.~\ref{contractions}).
We have not specified the flavor of the quark lines in the figure. In practice
where the flavor is taken into account, the contractions can be
more complicated according to the types of the currents (neutral or charged).
For the calculation with small momentum transfers, all the diagrams contribute and one needs to
combine them to have physical results. However,
for the case with large momentum and energy transfer as in the DIS, 
Figs.~\ref{ca}, \ref{cb} and \ref{cc} are dominated by the leading twist while 
Figs.~\ref{cd}, \ref{ce} and \ref{cf} are suppressed since they involve only high twists.
The respective leading twist parton degrees of freedom are classified by the first three diagrams, 
namely the valence and connected sea (CS) partons $q^{\rm v+cs}$ (Fig.~\ref{ca}), the CS anti-partons $\bar{q}^{\rm cs}$  (Fig.~\ref{cb})
and the disconnected sea (DS) partons and anti-partons $q^{\rm ds}+\bar{q}^{\rm ds}$ (Fig.~\ref{ce})~\cite{Liu:1993cv,Liu:1999ak}.
In our approach, they can be calculated separately which is a great feature especially for the
CS anti-partons that are responsible for the Gottfried sum rule violation~\cite{Liu:1993cv,Liu:1999ak,Liu:2012ch}.

After the Euclidean hadronic tensor is calculated, we need to convert it back to Minkowski space to obtain physical results.
Formally, the inverse Laplace transform fulfills this objective:
\begin{equation}
 W^M_{\mu\nu}(\vec{p},\vec{q},\nu)=\frac{1}{i}\int_{c-i\infty}^{c+i\infty}d\tau e^{\nu\tau}{W}^E_{\mu\nu}(\vec{p},\vec{q},\tau).
\end{equation}
However in practice, the Euclidean hadronic tensor is a function of Euclidean time, which is real, so that the integral in the inverse Laplace transform along the
imaginary time axis is not possible. Numerically, one can try to solve the inverse problem of the Laplace transform to determine an estimation of $W^M_{\mu\nu}$~\cite{Liu:2016djw,Liang:2017mye},
\begin{equation}
{W}^E_{\mu\nu}(\vec{p},\vec{q},\tau)=\int d\nu W^M_{\mu\nu}(\vec{p},\vec{q},\nu)e^{-\nu\tau}.
\end{equation}
Details about solving the inverse problem are discussed in the next section.

\section{\label{sec:Inverse}Solving the inverse problem}

A general form of the inverse problem reads
\begin{equation}
c(\tau_i)=\int k(\tau_i,\nu)\omega(\nu)d\nu,
\end{equation}
where $c(\tau_i)$ denotes discrete lattice data with finite number of points (usually $O(10)$),
$k(\tau_i,\nu)$ is the integral kernel that is a function of both $\tau_i$ and $\nu$, and $\omega(\nu)$
is the target function which is usually continuous with respect to $\nu$.
In principle, determining every detail of a totally unknown continuous function with finite input information is not possible; videlicet,
more than one solution can be found to match the input data. Numerically, we can discretize $\omega(\nu)$:
\begin{equation}
\label{INV1}
c(\tau_i)=\sum_j k(\tau_i,\nu_j)\omega(\nu_j)\Delta\nu_j;
\end{equation}
however, the number of $\nu_j$ one needs to reproduce the structures of $\omega(\nu)$ is, in many cases, much larger than
the number of input points, so the problem is still ill-posed. Nevertheless, many algorithms are available to extract the most probable solution
at a certain resolution.
Actually, this is a common problem, not only in physics, and the algorithms have been kept updated and improved. 

In this section, we will briefly introduce three methods of solving the inverse problem, discuss their features and use some mock data to
test their resolutions and robustness. 

\subsection{Backus-Gilbert method}
The Backus-Gilbert (BG) method~\cite{BG1,Backus123,Hansen:2017mnd}
utilizes the fact that the kernel functions can be linearly combined to approximate the Delta function, if they span a complete function basis
\begin{equation}
\sum_i a(\tau_j,\nu_0) k(\tau_i,\nu)\sim\delta(\nu-\nu_0),
\end{equation}
where $a(\tau_i,\nu_0)$ are the coefficients for the $i$'th kernel function at a specific point $\nu_0$,
which can be calculated by assuming a criterion of ``deltaness'' and solving the linear equations.
Having $a(\tau_i,\nu_0)$, the value of the target function at $\nu_0$ is
\begin{equation}
\sum_i a(\tau_i,\nu_0) c(\tau_i)\sim\int\delta(\nu-\nu_0)\omega(\nu)d\nu=\omega(\nu_0).
\end{equation}
It is worthwhile noting that the number of independent kernel functions is equal to the number of the discrete lattice data points so usually the function basis is far from
complete, leading to a coarse resolution.
In some sense, this is a feature instead of a disadvantage, since the broadened delta function (``regulated delta function'') 
can be treated as a smoothing procedure and ensures a well-defined
infinite volume limit~\cite{Hansen:2017mnd}. This is because the lattice spectrum is discrete and the volume correction for multi-particle states is not negligible. 
However in our case, we are only concerned 
about the inclusive contribution rather than the contribution from one particular state, the infinite volume limit is well-defined.
And in fact, it is nearly impossible to isolate the multi-particle-state contributions even with other methods of better resolution from the present day lattice data.
Another feature of BG is that,
unlike other Bayesian-type methods, it solves the problem point by point and dose not guarantee that the reconstructed result reproduces the input data~\cite{Kim:2018yhk}, so
careful checks are alway needed.

\subsection{Maximum Entropy method}
The Maximum Entropy (ME) method~\cite{rietsch1977maximum,Asakawa:2000tr} makes use of the Bayesian probability with 
prior information about the target function to find the most probable solution
\begin{equation}
P[\omega|D,\alpha, m]\propto\frac{1}{Z_SZ_L}e^{Q(\omega)},
\end{equation}
where $P[\omega|D,\alpha, m]$ denotes the probability that $\omega$ is the solution given lattice data  $D$, prior information $m$ and a hyper parameter $\alpha$.
$Q=\alpha S-L$ is a combination of the Shannon entropy
\begin{equation}
S=\sum_j \left[\omega(\nu_j)-m(\nu_j)-\omega(\nu_j){\rm log}\left(\frac{\omega(\nu_j)}{m(\nu_j)}\right)\right]\Delta\nu_j,
\end{equation}
which entails the constraint from the prior information, and the likelihood function
\begin{equation}
L=\frac{1}{2}\sum_{i,j}\left(c(\tau_i)-c^\omega(\tau_i)\right)C^{-1}_{ij}\left(c(\tau_j)-c^\omega(\tau_j)\right)
\end{equation}
with $c^\omega(\tau_i)$ being the correlator reconstructed using Eq.~(\ref{INV1}) and $C$ the covariance matrix,
which embodies the constraint from the data.
$m(\nu_j)$ is the default model (the prior information we plug in) and
$\alpha$ is the weight that balances the two constraints. If $\alpha$ is zero, ME reduces to the normal $\chi^2$ fit which has no 
unique solution for the inverse problem, since the number of parameters is larger than the number of input data. The uniqueness is guaranteed 
for finite $\alpha$ and the results of different $\alpha$'s are averaged in a range of $\alpha$ based on certain assumptions~\cite{Asakawa:2000tr}. 
Practically, the parameter space for finding the maximum probability $P[\omega|D,\alpha, m]$, i.e. the maximum value of $Q$, is reduced to
a smaller one instead of the whole $\nu$ space by employing singular value decomposition \cite{Bryan_1990, Asakawa:2000tr}, which makes
the maximum search easier while the resolution may be affected. Improved ME with an extended search space~\cite{Rothkopf:2011ef} is
also proposed and we will check whether it produces better results in our future study.

\subsection{Bayesian Reconstruction}
Bayesian Reconstruction (BR)~\cite{Burnier:2013nla} is an improved Bayesian method.
The Bayesian probability is
\begin{equation}
P[\omega|D,\alpha, m]\propto e^{Q'(\omega)},
\end{equation}
where 
$Q'=\alpha S'-L-\gamma(L-N_\tau)^2$ and 
\begin{equation}
S'=\sum_j \left[1-\frac{\omega(\nu_j)}{m(\nu_j)}+{\rm log}\left(\frac{\omega(\nu_j)}{m(\nu_j)}\right)\right]\Delta\nu_j,
\end{equation}
which is an alternative way to encode the constraint from the prior. $\gamma$ is a numerically large number such that 
the term $\gamma(L-N_\tau)^2$ helps to prevent over-fitting.
The hyper parameter $\alpha$ here is integrated over as
\begin{equation}
P[\omega|D,m]=\frac{P[D|\omega, I]}{P[D|m]}\int d\alpha P[\alpha|D,m],
\end{equation}
where $P[\alpha|D,m]$ is the probability of $\alpha$ in the present of $D$ and  $m$.
The search space of the maximum is also the whole parameter space which enhances the ability of finding the
real maximum while the price to pay is the need of high-precision architecture (e.g., 512-bit floating point numbers). 
For both Bayesian-type methods, ME and BR, the choice of the default model is
in principle arbitrary, but a reliable reconstruction should not depend on the default model. In all the following calculations of the paper, the 
default model is chosen to be a constant.

\begin{figure}[!h]
\includegraphics[width=.4\textwidth,page=1]{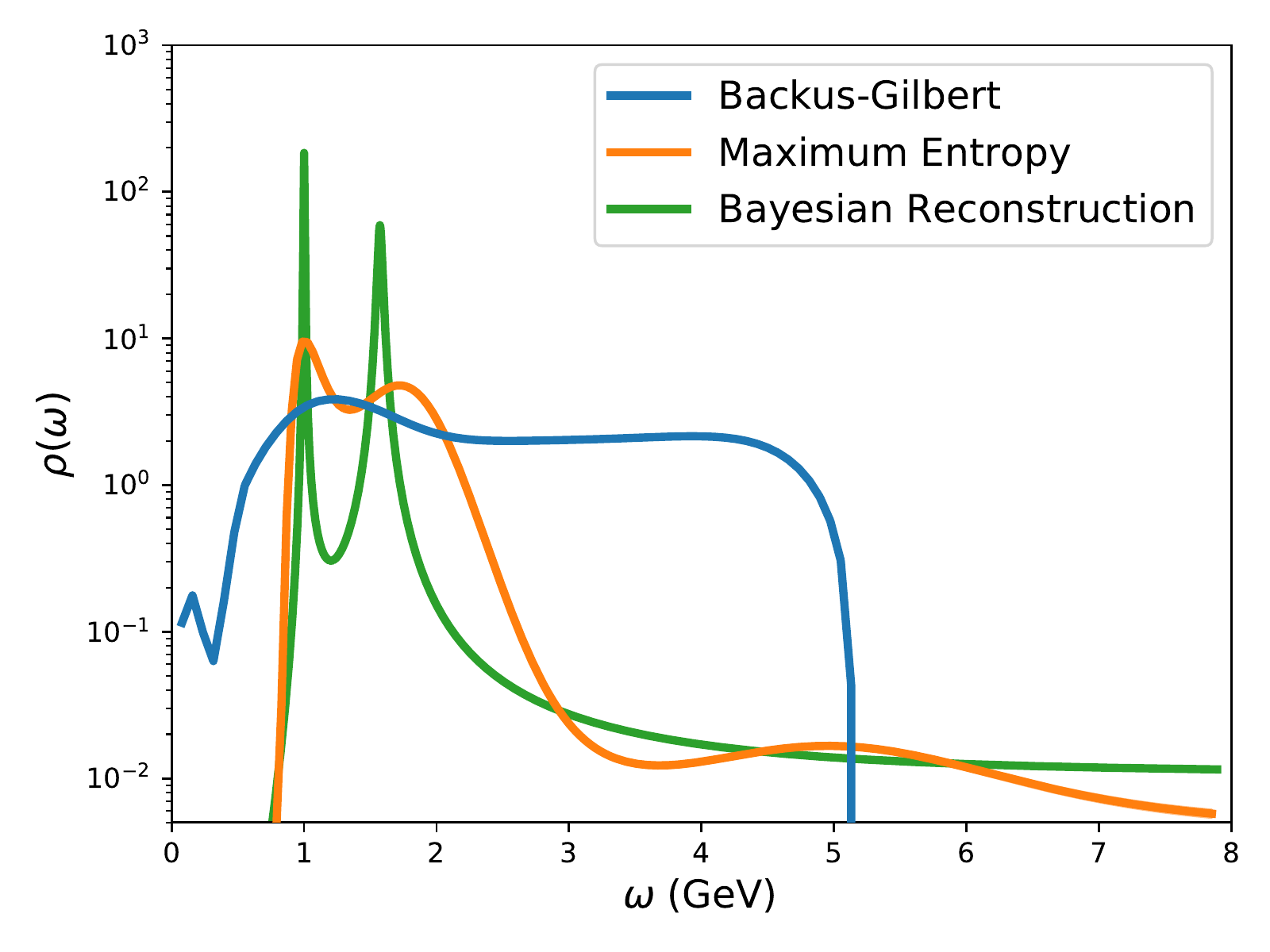}
\includegraphics[width=.4\textwidth,page=2]{comparison}
\caption{The results of the reconstruction of the first mock data set. The left panel shows the reconstructed spectral functions and the right panel shows the 
comparison between the original input data and the 2-point functions computed using the spectral functions.
Note that both the Maximum Entropy and Bayesian Reconstruction results coincide with the input data so the orange and green curves are not visible in the right panel. Bayesian Reconstruction has the best resolution in the test.\label{comparison}}
\end{figure}
To test the three methods, mock data are generated. 
The first set of mock data is a 2-point function with three states of mass 1.0, 1.5 and 1.8 GeV respectively and the spectral weights are all unity.
The lattice spacing is tuned to $0.1$ fm and the number of time slices available is 20. Noises are added by assuming normal distributions around the
central values and the signal-to-noise ratio is set to be 100.
This is to check the resolution for peak structures.
The results of the reconstruction can be found in Fig.~\ref{comparison}. The left panel shows the reconstructed spectral functions and the right panel shows the 
comparison between the original input data and the 2-point functions computed using the spectral functions.
For the spectral functions we see that the result of BG shows basically no peak structures, reflecting its poor resolution for this setup.
The result of ME is much better as two peaks around 1 and 2 GeV are clearly seen. Considering the broad widths of the peaks,
it is consistent with the input data.
BR gives the best reconstruction in this case as two sharp peaks appear at $\sim1$ and $\sim1.6$ GeV. Although the second and the third 
states are not separated, the resolution of BR is much better. 
From the right panel we see that the regenerated 2-point functions from the spectral functions of ME and BR are well consistent with the input data,
but for the BG case, differences occur at large $t$. This exhibits the fact that, as pointed out before, 
BG does not guarantee that the reconstructed result reproduces the input data.

\begin{figure}[!h]
\includegraphics[width=.4\textwidth,page=1]{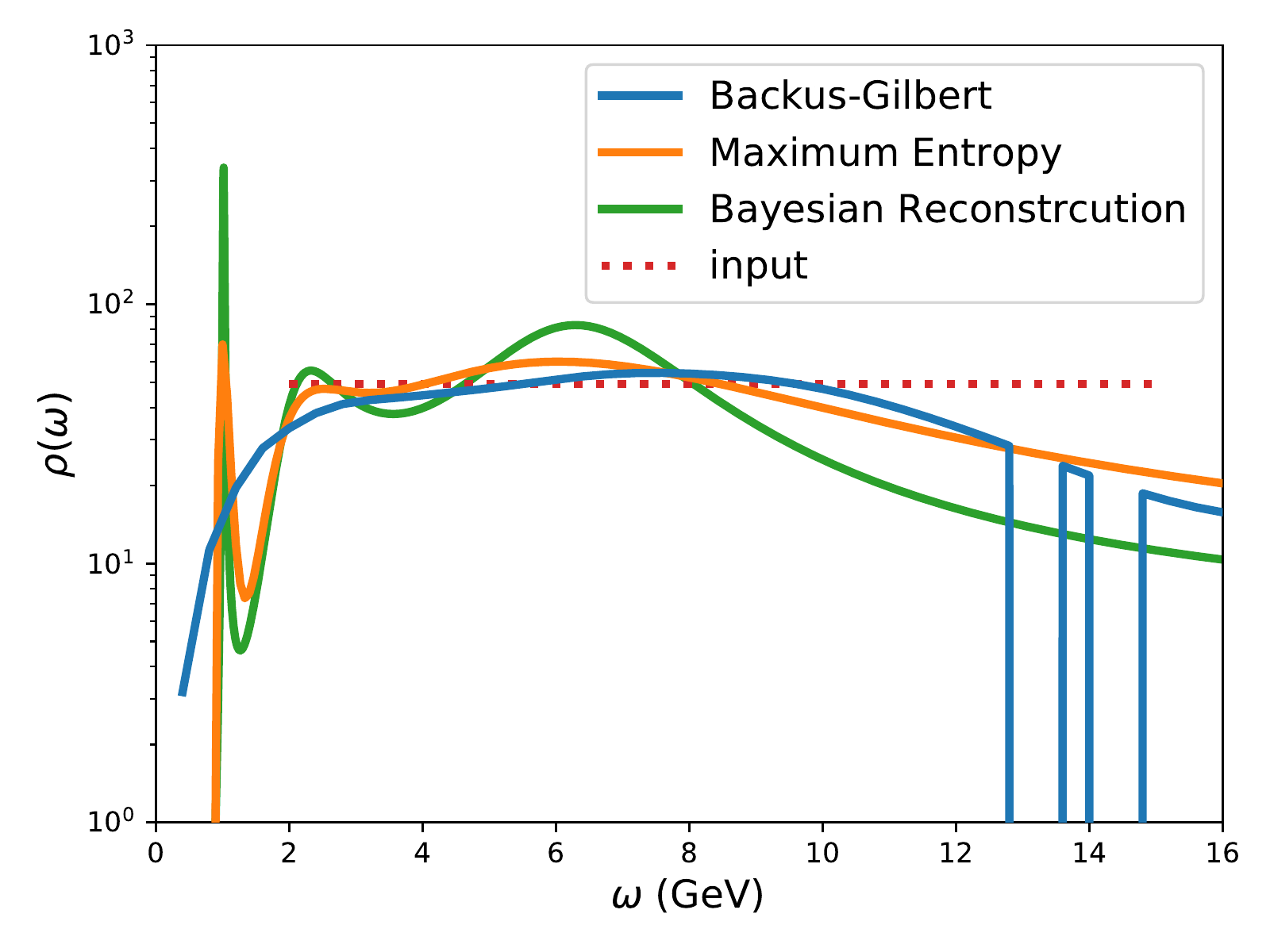}
\includegraphics[width=.4\textwidth,page=2]{comparison2}
\caption{The same as Fig.~\ref{comparison} but for the second mock data set. Again, both the Maximum Entropy and Bayesian Reconstruction results coincide with the input data so the orange and green curves are not visible in the right panel. \label{comparison2}}
\end{figure}
The second set of mock data is a 2-point function with two isolated states of mass 1.0, 1.5 and a dense spectrum (simulating the continuous spectrum) 
from 2 GeV to 15 GeV. This is like the energy dependence of the neutrino-nucleon scattering cross section.
The lattice spacing is set to $0.02$ fm in this case and the number of time slices is 100. The signal-to-noise ratio is set to be 100 too.
This is to check the case of a flat spectral function up to higher energies.
The results of the this test are shown in Fig.~\ref{comparison2}. 
Similarly, the left panel shows the reconstructed spectral functions and the right panel shows the 
comparison between the original input data and the 2-point functions recomputed using the spectral functions.
From the left panel we see that both ME and BR reconstruct a peak at $\sim 1$ GeV (the BR one is much sharper) while BG shows no peak in that 
energy region. This agrees with what we observed in the previous test. 
However, in the region 2 to 8 GeV, BG presents more consistent results with 
the input (the red dashed line) while both ME and
BR show unphysical oscillations.
These unphysical oscillations
are called ``ringing'' and are artifacts of the reconstruction~\cite{Kim:2018yhk}.
The ringing of the 
ME method seems weaker, which is due to the fact that SVD contains an additional smoothing that
suppresses the ringing but also leads to the significantly larger width of
peak structures.
Both the BR and ME methods suffer from this disadvantage, so they are not the optimal method for reconstruction of flat spectral functions. Actually, there is an update of BR aiming
to address this problem~\cite{Fischer:2017kbq,Kim:2018yhk} and we will also try to test this in our future study.
Similar to the first case, the right panel of Fig.~\ref{comparison2} 
shows that the regenerated 2-point functions from the spectral functions of ME and BR are well consistent while that of BG
is not.

Solving the inverse problem is the most challenging part of our calculation. More detailed studies on
how the lattice spacing, the error of the correlation functions and the number of time slices affect the reconstruction and
more inverse methods regarding this physics problem are needed. 
Recently, several inverse algorithms have been applied to evaluate the efficiency of obtaining $x$-dependent PDF's from 
mock Euclidean time correlators~\cite{Karpie:2019eiq}.
From the above two tests we know that BR has the best resolution for peak structures while BG and ME are more stable for flat
spectral functions.
Knowing the different methods' advantages, one can combine them to resolve different parts of the
spectrum, e.g., BR for the sharp peak structures and ME for the flat region.

\section{\label{sec:results}Preliminary results}
\subsection{Elastic case}

Having discussed how to solve the inverse problem, we now apply the algorithms to realistic lattice data.
The first example is to check the vector charge. This is a calculation that is relevant to neutrino-nucleon scattering and also
serves as a benchmark of the whole approach. The calculation is done on RBC/UKQCD domain wall lattice 32Ifine~\cite{Blum:2014tka} with
clover fermions as valence quarks. The configuration is preprocessed by HYP smearing and 
the tadpole improved clover coefficient $C_{\rm sw}=1.033$ is used to generate the clover term of the clover action
for the valence quarks. The pion mass is tuned to be close to the unitary point $\sim371$ MeV.
The lattice spacing
is about 0.06 fm which we expect to be fine enough such that the inverse algorithms can give reasonably reliable results.

\begin{figure}[!h]
\begin{subfigure}[b]{0.3\textwidth}
\includegraphics[width=\textwidth,page=1]{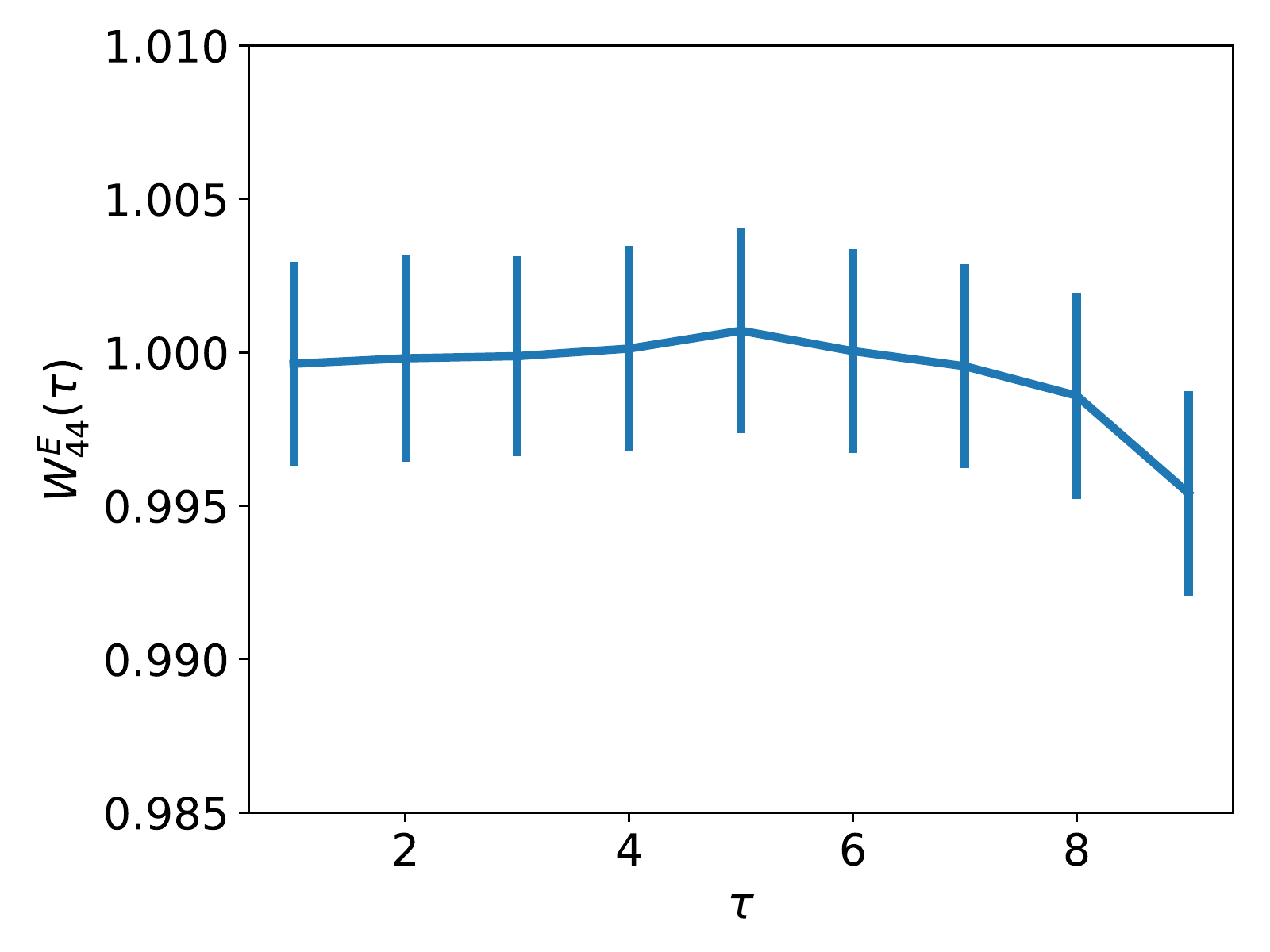}
\caption{\label{charge1}}
\end{subfigure}
\begin{subfigure}[b]{0.3\textwidth}
\includegraphics[width=\textwidth,page=1]{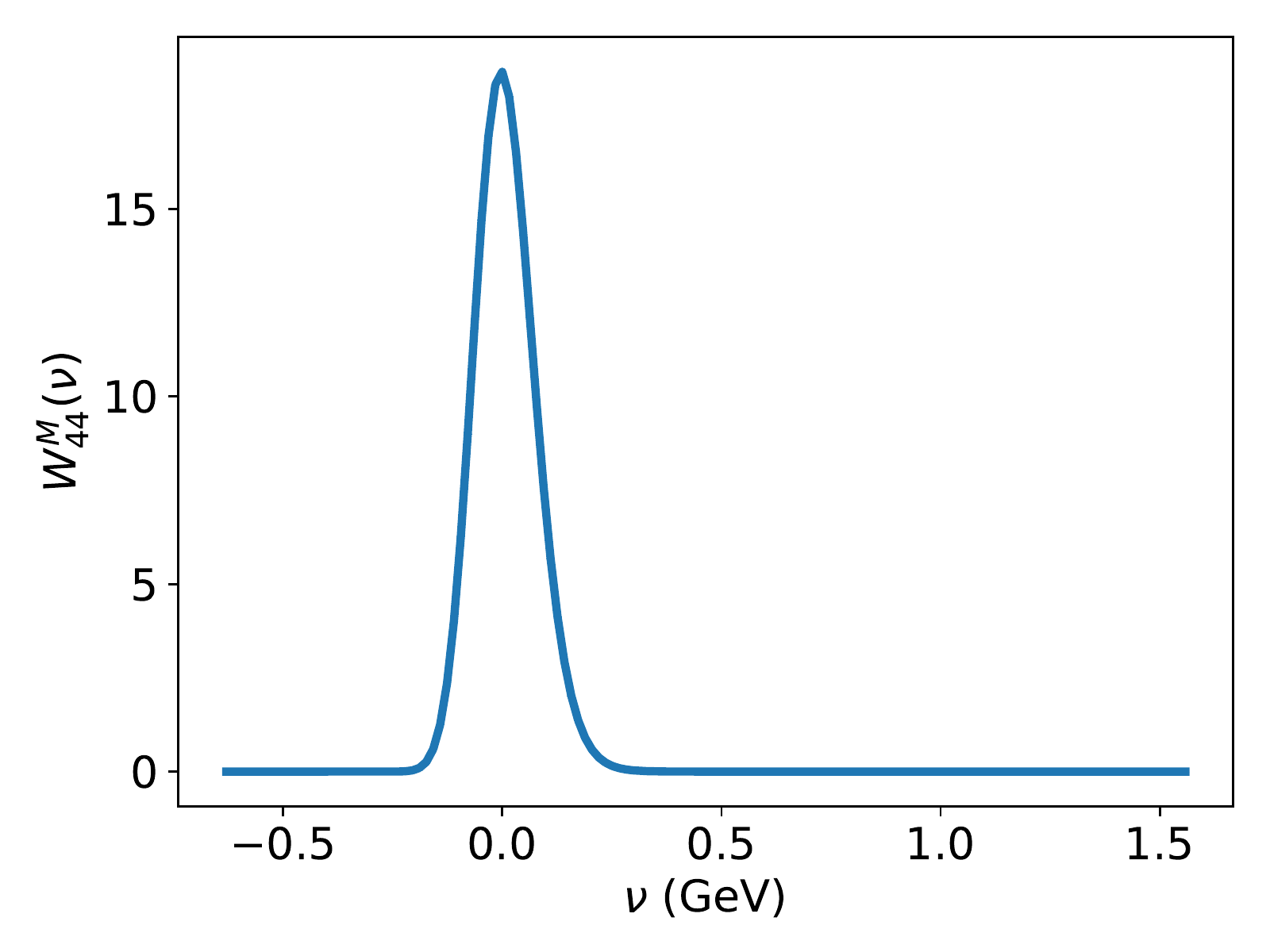}
\caption{\label{charge2}}
\end{subfigure}
\begin{subfigure}[b]{0.3\textwidth}
\includegraphics[width=\textwidth,page=1]{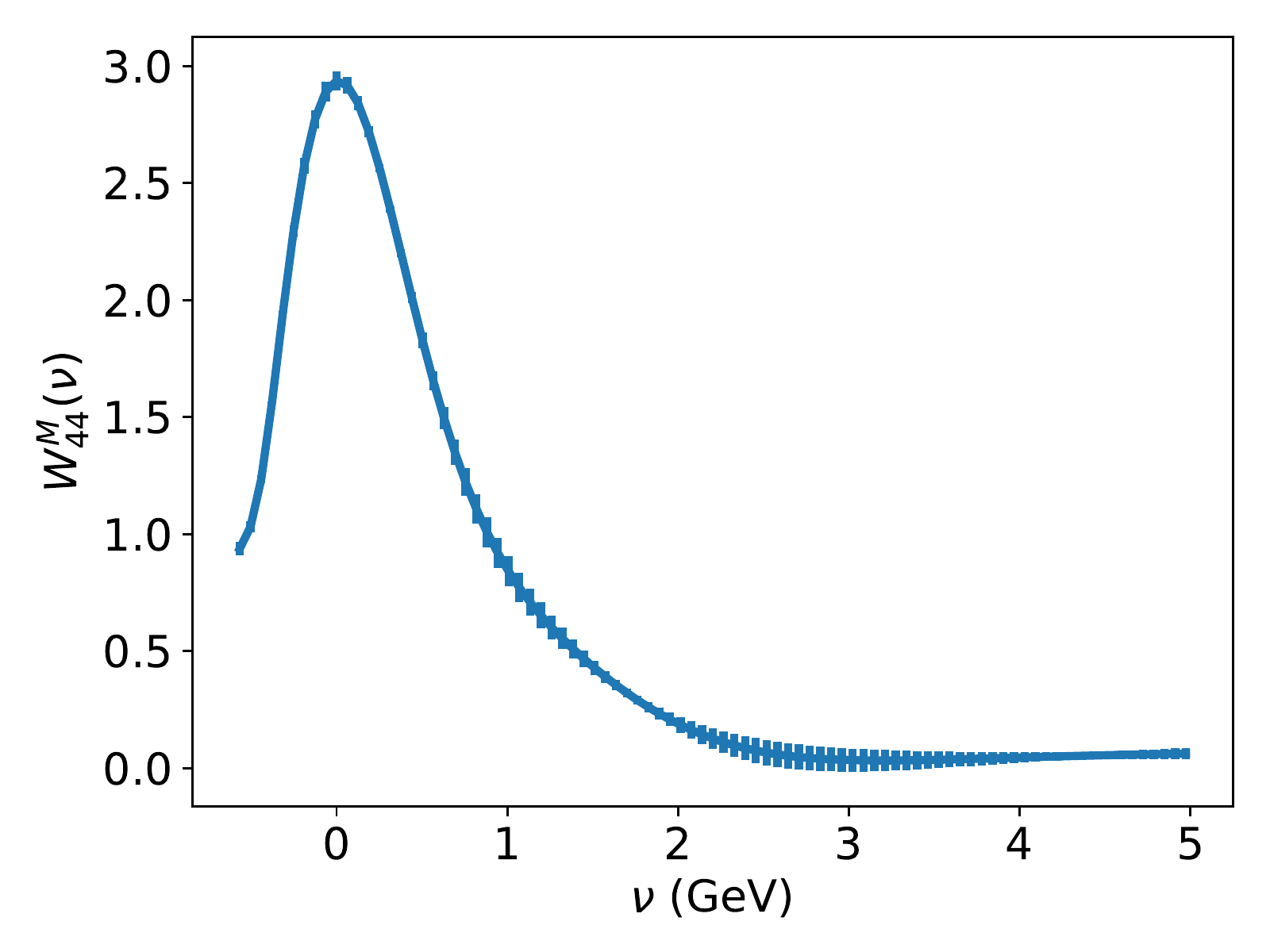}
\caption{\label{charge3}}
\end{subfigure}
\caption{The results of the elastic case. Fig.~\ref{charge1}: the Euclidean hadronic tensor ${W}^E_{44}$ as a function of $\tau$.
Fig.~\ref{charge2}: the Minkowski hadronic tensor ${W}^M_{44}$ as a function of energy transfer $\nu$ from the ME method.
Fig.~\ref{charge3}: the Minkowski hadronic tensor as a function of $\nu$ from the BG method.
}
\end{figure}

For this case, we choose $\mu=\nu=4$ and $\vec{p}=\vec{q}=0$ for ${W}^E_{\mu\nu}$. So Eq.~(\ref{eq:EHT}) becomes
\begin{equation}
{W}^E_{44}(\vec{0},\vec{0},\tau)\sim\sum_n\langle p,s|\bar{\psi}\gamma_4\psi| n \rangle\langle n|\bar{\psi}\gamma_4\psi|p,s\rangle e^{-(E_n-M_p)\tau}.
\end{equation}
For simplicity, the two currents are both inserted on the $d$ quark line so only Fig.~\ref{ca} contributes. 
And for $\tau\gg 0$ only the ground state survives, so ${W}^E_{4,4}=g_V^2=1$, given proper normalization factor $Z_V$.
Two sequential propagators are used for constructing the 4-point function with one starting from $t_0$ through $t_1$ to $t_2$ and the other starting from $t_0$ through $t_f$ to $t_2$ (Fig.~\ref{ca}).
Therefore, for each calculation, the source point $t_0$ and the two sequential points $t_1$ and $t_f$ are fixed while all values of $t_2$ are available.
In this particular example, we choose $t_0=0$, $t_f=15$ and $t_1=5$ in lattice unit so $t_2$ should be in the range of $[6, 14]$ to exclude the contact points and the corresponding $\tau=t_2-t_1$ is from $1$ to $9$.
The result of ${W}^E_{44}(\tau)$ is plotted in Fig.~\ref{charge1}. It shows that within errors, ${W}^E_{4,4}(\tau)$ is indeed a constant of value $1$. The drop at $\tau=9$ is likely due to the fact 
that it is too close to the nucleon sink. The errors are around $0.4\%$ and the number of configurations used is $100$.

To convert the results to the Minkowski space, the ME and BG methods are employed. Actually, the exact form 
of hadronic tensor for elastic scatterings~\cite{ChengTa-Pei1984Gtoe} is
\begin{equation}
W^M_{44}({q}^2, \nu)=\delta(q^2+2m_p\nu)\frac{2m_p}{1-q^2/{4m_p}^2}\left(G_E^2(q^2)-\frac{q^2}{4M_N^2}G_M^2(q^2)\right).
\end{equation}
In our case $W^M_{44}(\nu)\overset{q^2=0}{=}\delta({\nu}) G^2_E(0)=\delta(\nu)$. This is easy to understand since the spectral function should be
a Dirac delta function at $\delta(\nu-\nu_0)$ when the Euclidean correlator is a single exponential $\sim e^{-\nu_0t}$ and here the constant is a special case of an exponential
with $\nu_0=0$.

The converted results of $W^M_{44}(\nu)$ using ME and BG are plotted in Fig.~\ref{charge2} and Fig.~\ref{charge3}, respectively.
They both give a peak around $\nu=0$ and in this sense the results are consistent with the theoretical prediction of $\delta(v)$. However, similar to the cases of the mock data,
ME shows much better resolution than BG.
Another problem of the result of BG is that it is not symmetric about $\nu=0$, which is because BG has difficulties in resolving the results of negative $\nu$.
An important check is that the area under the peaks should be $g_V^2=1$. The values of numerical integral of the results from BG and ME are 
1.18(6) and 1.001(7) respectively. Again, ME shows a more precise result. 
Although it is not necessary and cumbersome to calculate the vector charge by
constructing the 4-point function and solving the inverse problem,
it nevertheless shows the feasibility of our approach. 
The vector charge can be obtained reliably. For more complicated cases such as non-zero momentum transfers or charged currents,
this approach may show its advantages and provide the inclusive contribution of all intermediate states.

\subsection{Non-zero nucleon momentum and momentum transfer}

As pointed out in the introduction, another important motivation of calculating the hadronic tensor is to have the lattice results of structure functions in the DIS region which can be
used together with experimental inputs to better pin down the parton distribution functions. To this end, we need to have large momentum transfers to
make the scattering ``deep'' enough to access the parton degrees of freedom. Meanwhile, the momentum of the external proton cannot be
too small (and it is better to be in the opposite direction of the momentum transfer) if one wants to reach small $x$ (e.g. $\sim0.1$). 
For this calculation, we use an anisotropic clover lattice~\cite{Lin:2008pr} with $a_t\sim0.035$ fm.
The pion mass is about 380 MeV and the momentum unit is $\frac{2\pi}{L_s}\sim0.42$ GeV.
The reason we switch to this lattice is that the signal-to-noise ratio will be much worse than the previous elastic case when the momentum transfer or
the nucleon momentum is large. Thus, having more data points in the $t$ direction helps the inverse algorithms to have more stable results.

The detailed kinetic setup is listed in Table.~\ref{TB.mom}. We choose $\vec{p}=(0,3,3)$ and $\vec{q}=(0,-6,-6)$ in lattice unit and $\mu=\nu=1$,
such that only the $F_1$ structure function survives; thus $W^M_{11}=F_1(x,Q^2)$.
Since the energy transfer $\nu$ is not fixed by the lattice 3-dimensional Fourier transform, we can choose a range of $\nu\in[2.96, 3.68]$ GeV such that the corresponding  
$Q^2$ is in a range of 2 to 4 GeV$^2$. The Bjorken $x$ that can be accessed is between 0.07 and 0.16 for this setup. An interesting point of 
this setup is that $\vec{p}+\vec{q}=-\vec{p}$, therefore the energy of the lowest intermediate state $E_{n=0} = E_p$ and for large enough $\tau$,
${W}^E_{11}\propto e^{-(E_{n=0}-E_p)\tau}$ is a constant.

\begin{table}[!h]
\centering
\begin{tabular}{cccccccc}
\hline
$\vec{p}$ (${2\pi}/{L_s}$) & $\vec{q}$ (${2\pi}/{L_s}$) & $E_p$ (GeV) & $E_{n=0}$ (GeV) & $|\vec{q}|$ (GeV) & $\nu$ (GeV) & $Q^2$ (GeV$^2$) & $x$ \\ \hline
(0,3,3) & (0,-6,-6) & 2.15 & 2.15 & 3.57 & [2.96, 3.68] & [4, 2] & [0.16, 0.07] \\ \hline 
\end{tabular}
\caption{The kinematic setup of nucleon momentum $\vec{p}$, three-momentum transfer $\vec{q}$, proton energy $E_p$,
the energy of the lowest intermediate state $E_{n=0}$, the modulus of the three-momentum transfer $|\vec{q}|$,  the range of 
energy transfer $\nu$, the range of four-momentum transfer $Q^2$ and the corresponding Bjorken $x$.
\label{TB.mom}}
\end{table}

\begin{figure}[!h]
\includegraphics[width=.4\textwidth,page=1]{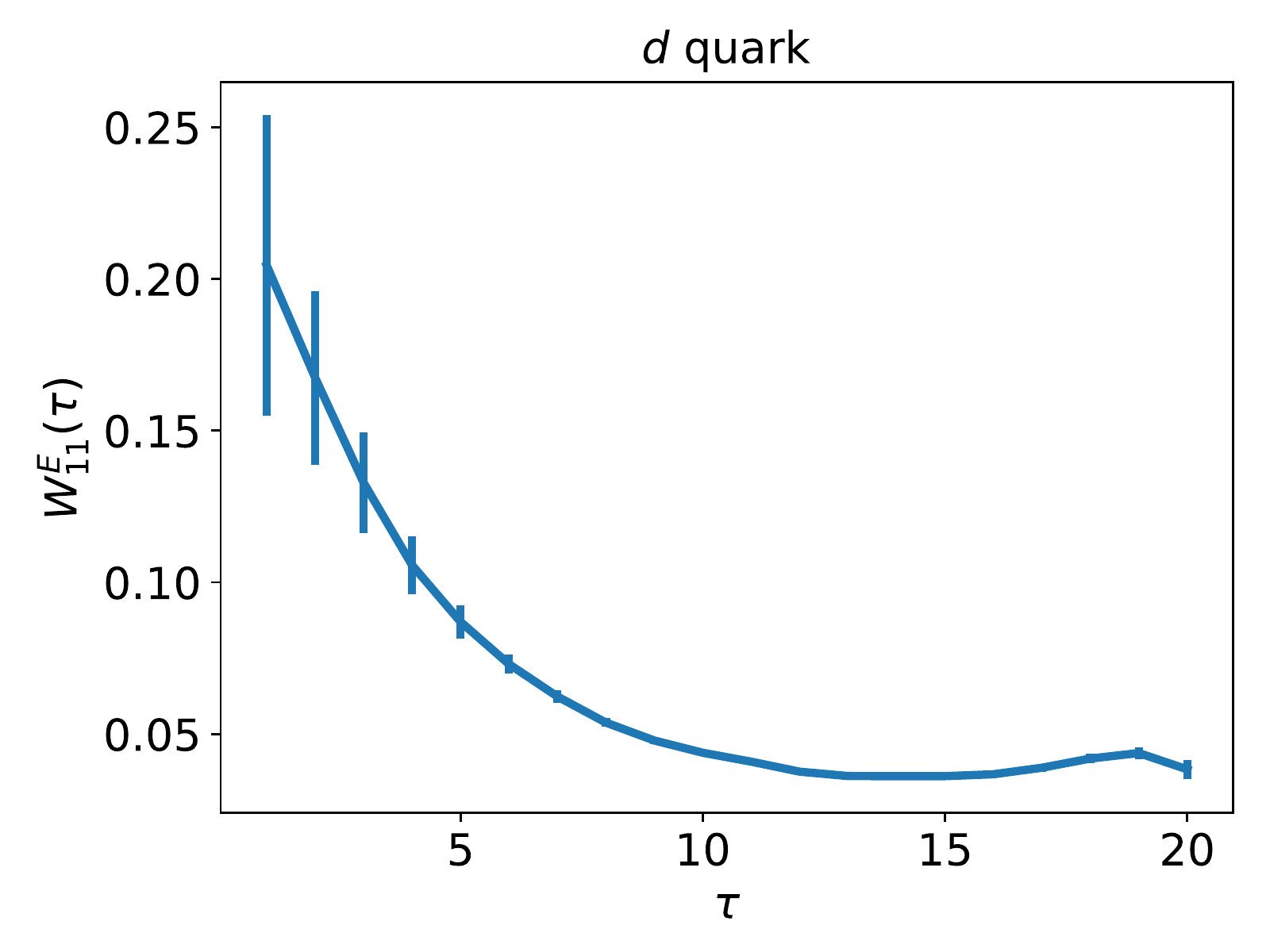}
\includegraphics[width=.4\textwidth,page=1]{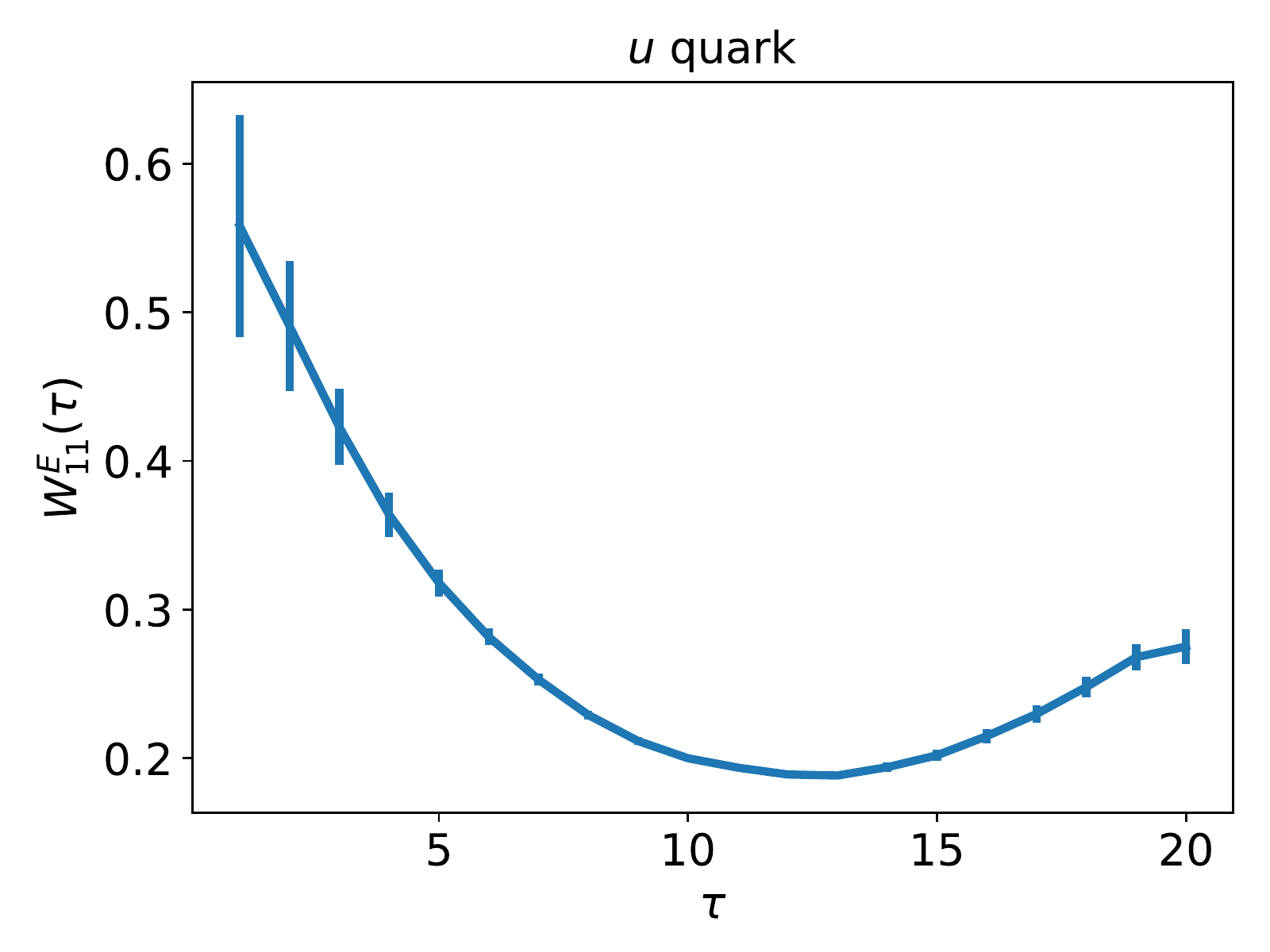}
\caption{The Euclidean hadronic tensor ${W}^E_{11}$ as a function of $\tau$ for both $d$ and $u$ quarks. $\vec{p}=(0,3,3)$ and $\vec{q}=(0,-6,-6)$ in lattice unit in this case. \label{E2}}
\end{figure}

The results of the Euclidean hadronic tensor ${W}^E_{11}(\tau)$ for both $d$ and $u$ quarks are shown in Fig.~\ref{E2}.
In this calculation, we set $t_0=0$, $t_1=8$ and $t_f=28$ in lattice unit and still we focus on the diagram in Fig.~\ref{ca} only for the moment.
The figures show that for small $\tau$, the Euclidean hadronic tensor ${W}^E_{11}(\tau)$ decays exponentially since the energy of the
intermediate states $E_n$ are larger than $E_p$. For larger $\tau$, it gets flatter for the $d$ quark case (left panel) which is consistent with what we expect, i.e. $E_{n=0}=E_p$.
The tail of the Euclidean hadronic tensor for the $u$ quark case (right panel) goes up after $\tau\sim15$, which we believe is due to the contamination of the sink nucleon excited states.

\begin{figure}[!h]
\includegraphics[width=.4\textwidth,page=1]{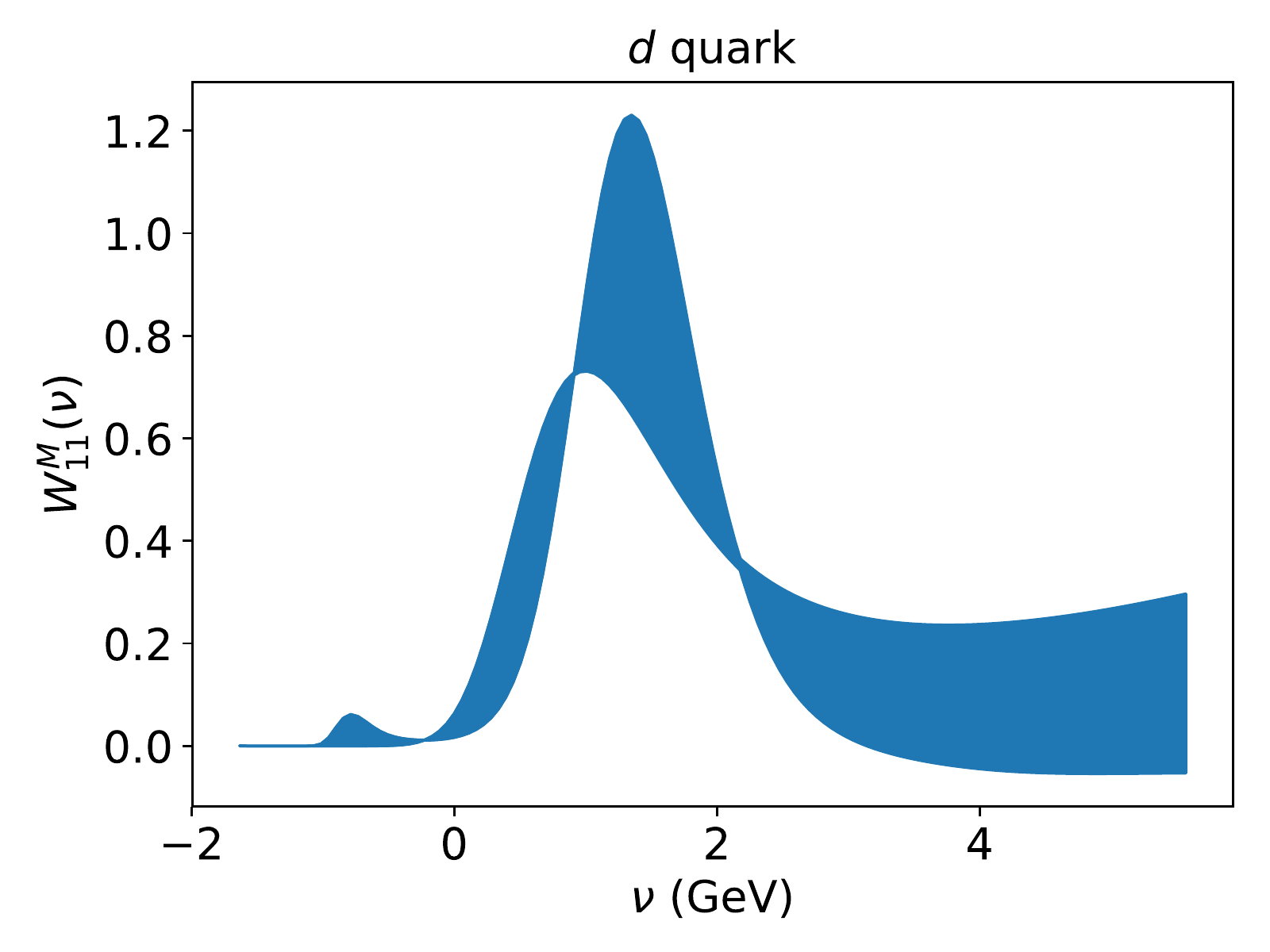}
\includegraphics[width=.4\textwidth,page=1]{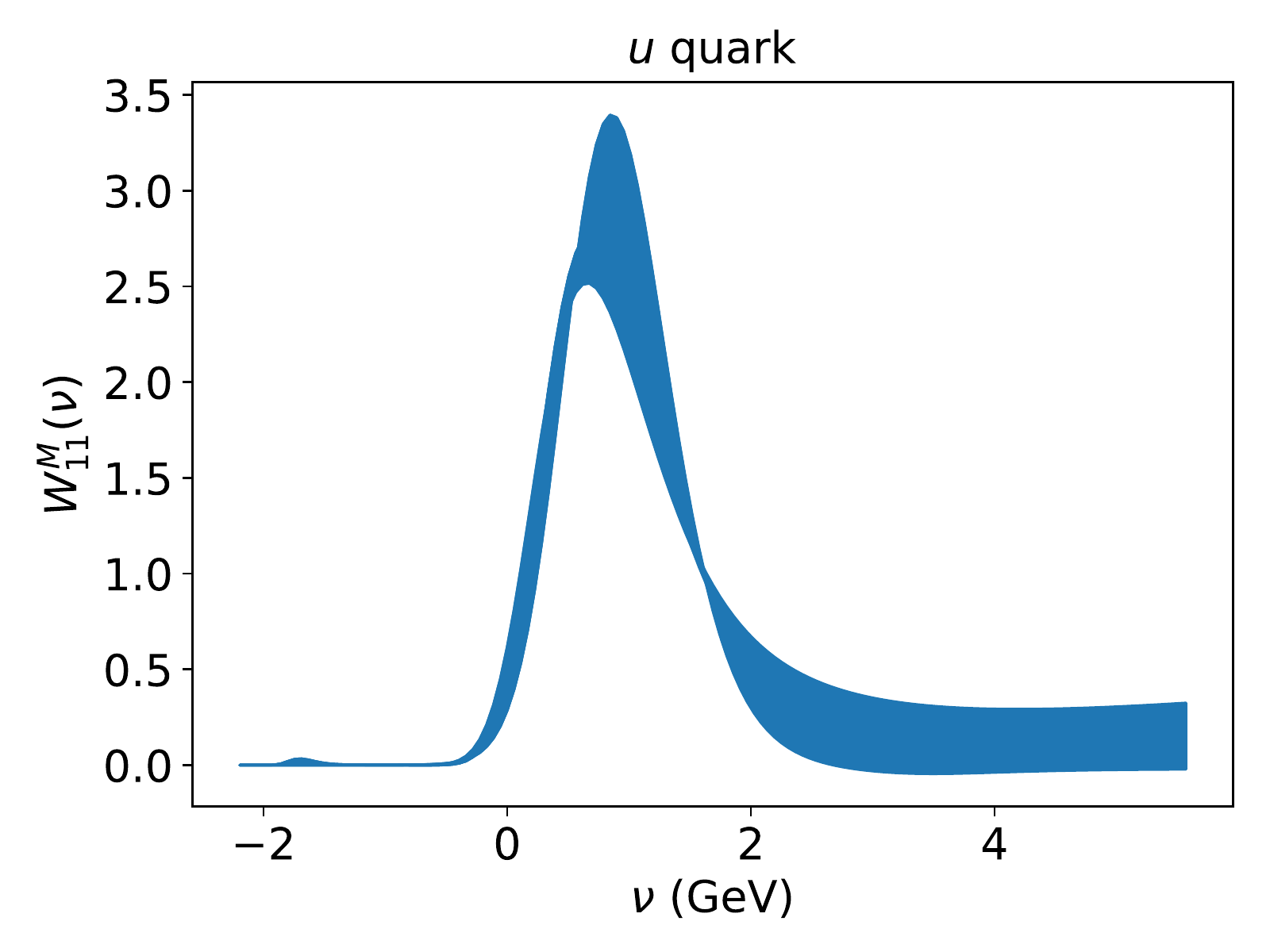}
\caption{The Minkowski hadronic tensor ${W}^M_{11}$ as a function of energy transfer $\nu$ reconstructed from the ME method for both $d$ and $u$ quarks.
The error bands show the difference introduced by different ME parameters while no statistical errors are included. At some certain points, the error seems tiny 
which just indicates that different ME parameters result in similar results.\label{M2}}
\end{figure}

\begin{figure}[!h]
\includegraphics[width=.4\textwidth,page=1]{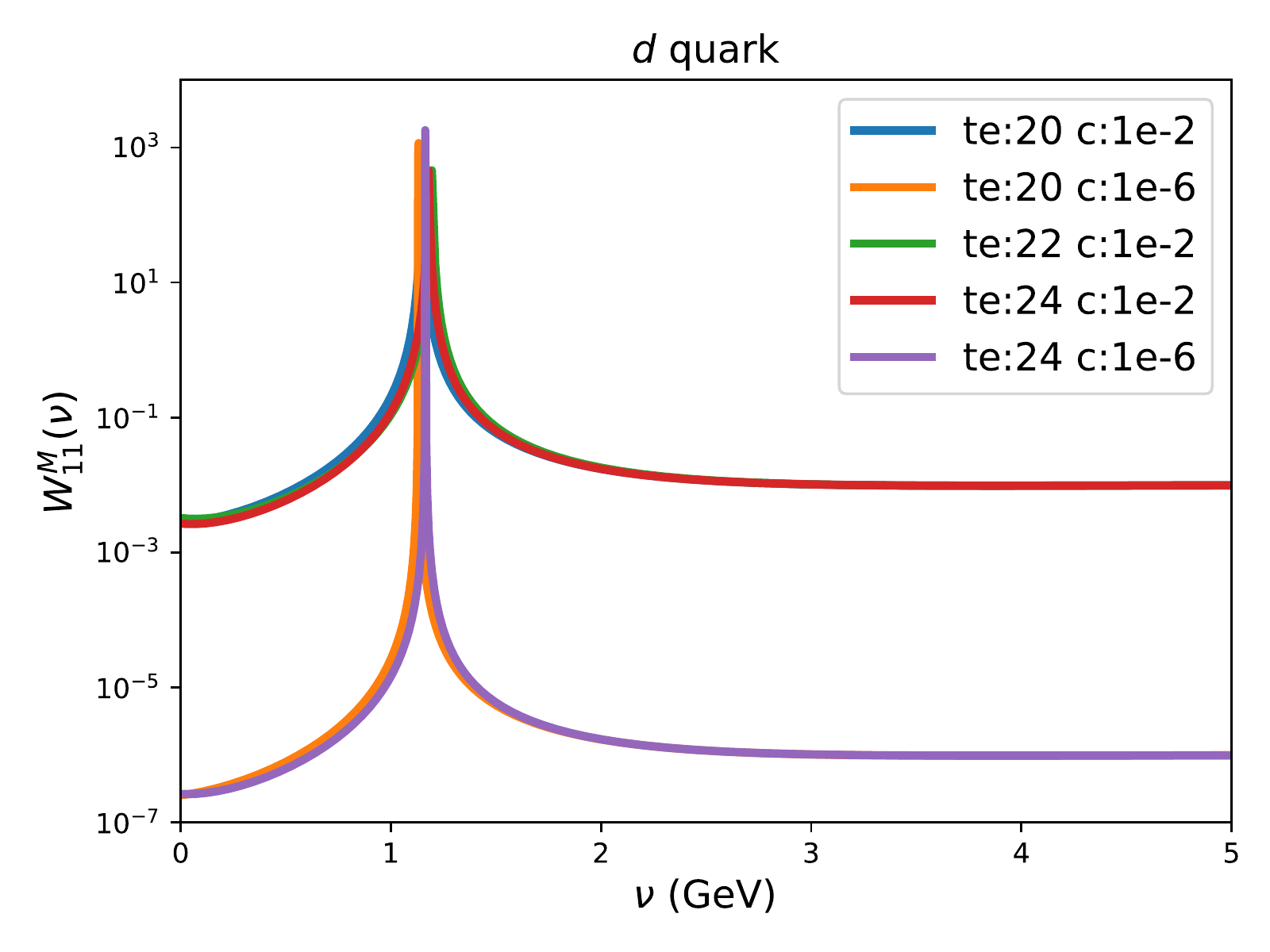}
\includegraphics[width=.4\textwidth,page=1]{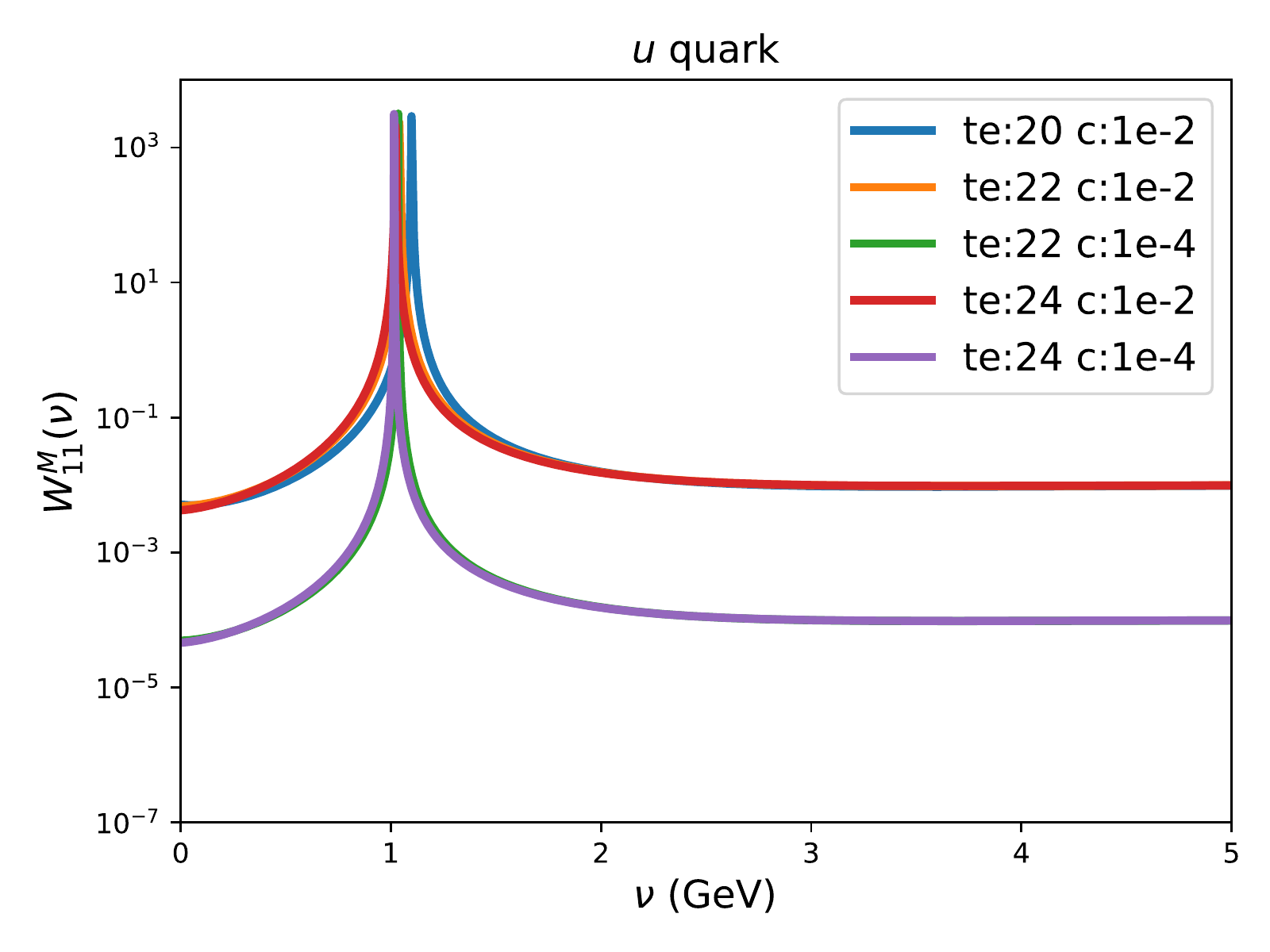}
\caption{The Minkowski hadronic tensor ${W}^M_{11}$ as a function of energy transfer $\nu$ reconstructed from the BR method for both  $d$ and $u$ quarks.\label{M3}
In the label, $te$ denotes the end point of $t_2$ we use for the reconstruction and $c$ is for the value of constant default model. After $\sim$ 2 GeV, the default models
dominate the results.}
\end{figure}

Similarly, we need to solve the inverse problem to obtain results in Minkowski space.
The results from the ME method are shown in Fig.~\ref{M2}. The error bands are mainly from
the average of different default models. 
The behaviors of $d$ and $u$ quarks are similar.
We do not observe a peak around the elastic point $\nu=0$
which is because at this point the hadronic tensor is the square of the electromagnetic form factor 
and the form factor for elastic scattering is highly suppressed.
Taking a dipole form for the form factor with $Q^2 =12.7~\rm{GeV}^2$ in this case, 
the hadronic tensor is suppressed by a factor of $(1 + Q^2/0.71)^{-4}\sim 10^{-5}$ as compared to the charge at $Q^2 = 0$.
We do observe a broad structure shows at about 1 GeV, which should be the combined contribution of nucleon resonances and possibly SIS. 
As discussed above, the preferred $\nu$ range that can lead us to the parton structure functions is from 2.96 GeV to 3.68 GeV, however, 
our results show that it is basically zero within error in that region.
To check whether this is a resolution issue of ME, we 
also use the BR method that shows better resolution for discrete structures in our mock data test to handle the same data. The results are shown in Fig.~\ref{M3}.
And this time, to show exactly the effect of different default models, we plot the results with different default models
separately in log scale.
We also check the effect of including or excluding the data points of large $\tau$ since they can have large excited-state contaminations.
This time, except that the peaks around 1 GeV are much shaper than the ME case, the basic conclusion we learn is 
the same. No elastic contribution shows at $\nu=0$ and the only structure is around 1 GeV. When the energy transfer
goes above, say, 2 GeV, the reconstructed results approach to those of the default model values which means the data have no constraint in
that region.

From these two examples, we see that the elastic and the resonance contributions, where the energy transfer 
is not too large, can be extracted for the current setup. However, it seems that there are no contribution in the DIS region.
We will discuss the possible reason and solution in the next section.

      It is suggested by \mbox{A.J. Chambers {\it et al.}~\cite{Chambers:2017dov}} that the structure functions can be extracted through the time-ordered current-current correlator on the lattice through the operator product expansion of the forward Compton amplitude. This is carried out by setting $\nu = 0$ in the integral in Eq.~(\ref{Laplace}) to avoid
the divergence when $\nu - (E_n - E_p) > 0$ as discussed in Sec.~\ref{sec:form} after Eq.~(\ref{Laplace2}). However, this leads 
$W_{\mu\nu}' (p,q,T)$ in Eq.~(\ref{Laplace2}) (considered to be the Compton amplitude $T_{\mu\nu}(p,q)$ in 
~\cite{Chambers:2017dov}) to 
\begin{eqnarray}  \label{Compton}
&W'_{\mu\nu}(T)&= \int d^{3}\vec{z}\frac{e^{i\vec{q}\cdot\vec{z}}}{4\pi} [\langle p,s|J_{\mu}^{\dagger}(\vec{z})|0\rangle\langle 0|J_{\nu}(0)|p,s\rangle T \nonumber \\
&&\!\!\!\!\!\!\!+\sum_{n\geq 1}\frac{e^{\left(\nu-(E_{n}-E_{p})\right)T}-1}{\nu-(E_{n}-E_{p})}\langle p,s|J_{\mu}^{\dagger}(\vec{z})|n\rangle\langle n|J_{\nu}(0)|p,s\rangle],
\end{eqnarray}
where the state label $0$ is the nucleon state with momentum $\vec{p} + \vec{q}$ so that the first term is the elastic scattering
which diverges as $T$ and reflects the elastic scattering pole in Eq.~(\ref{Laplace2}). As $T \rightarrow \infty$, the $n \geq 1$ state contributions are suppressed by $1/T$. 
However, as shown in Fig.~\ref{M2} when $T$ is finite (0.7 fm in our case), the excited states including nucleon resonances and those in the SIS and DIS regions, all contribute. 
When $Q^2$ is large (e.g., 12.7 $\rm{GeV}^2$ in this case), the hadronic tensor for the elastic scattering is highly suppressed (by a factor of $\sim 10^{-5}$). Whereas, the resonance contribution around $1-2$ GeV at  $Q^2 \sim 9 - 11\, \rm{GeV}^2$ is much larger as shown in Fig.~\ref{M2}. To estimate
how large a $\nu$ is needed for DIS, we can look at $W$, the total invariant mass of the hadronic final state
\begin{equation}
W^2 = (q+p)^2 =m_p^2 -Q^2 + 2 m_p\, \nu.
\end{equation}
The global fitting of PDF usually take a cut with \mbox{$W^2 > 10\, \rm{GeV}^2$.} When we take $Q^2 = 4\, \rm{GeV}^2$, this gives $\nu > 6.5$ GeV. Therefore, taking $\nu = 0$ in Eq.~(\ref{Laplace}) will not yield PDF in the DIS region which needs both $Q^2$ and $\nu$ to be large.

\section{\label{sec:discussion}Discussion and Summary}

\begin{figure}[!h]
\includegraphics[width=.4\textwidth,page=1]{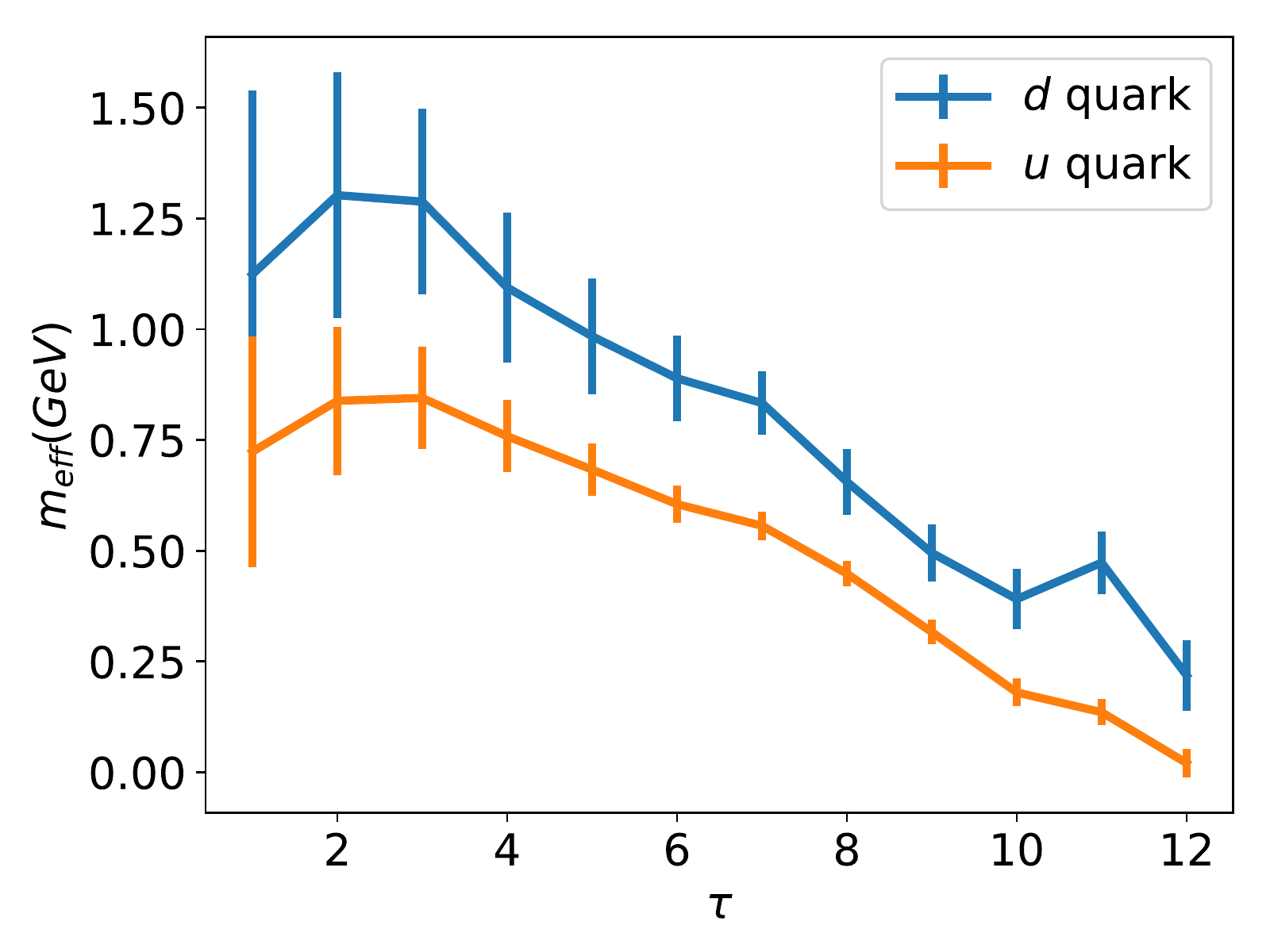}
\includegraphics[width=.4\textwidth,page=1]{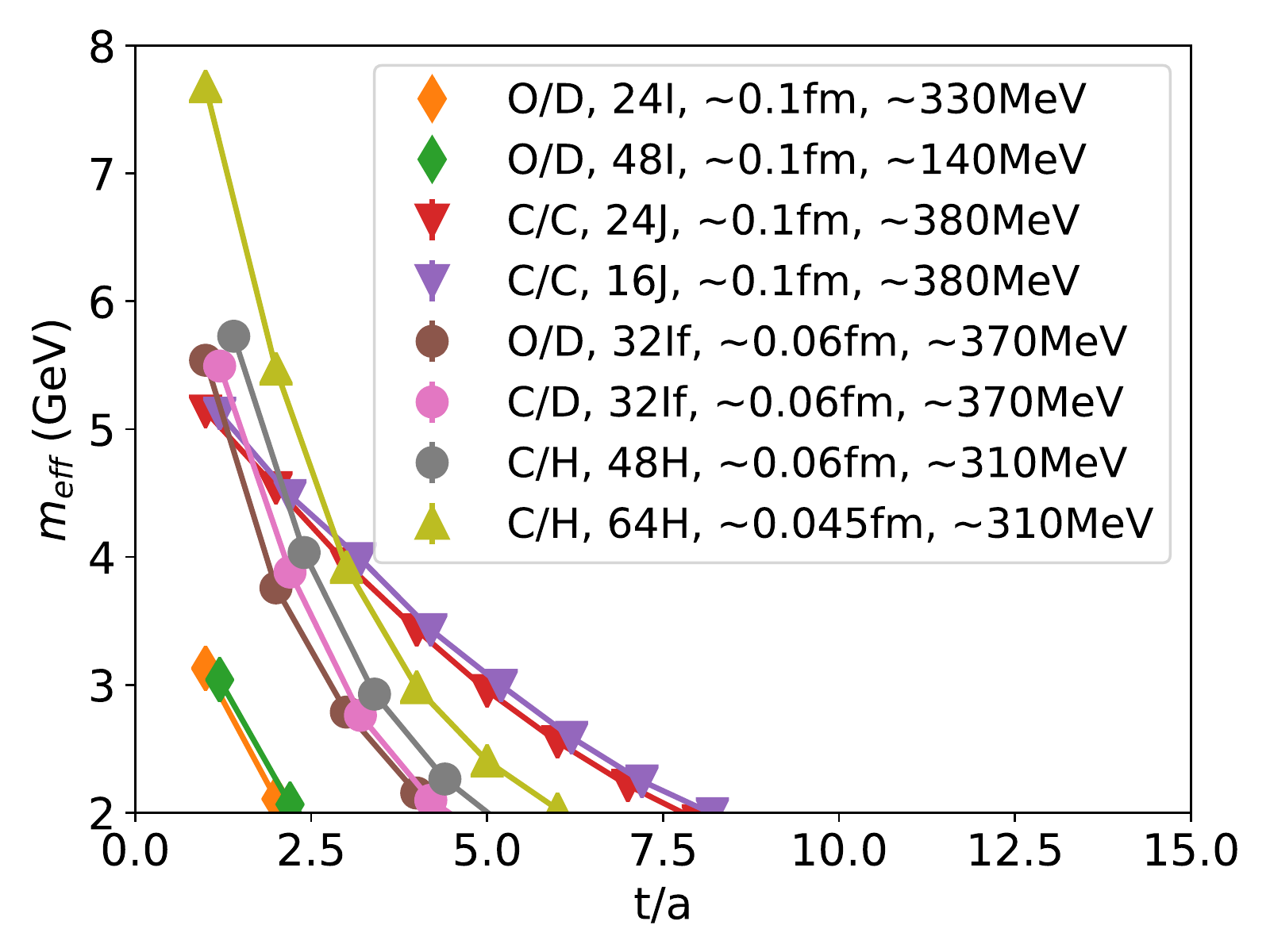}
\caption{The effective mass plot for the Euclidean hadronic tensor (left panel) and for the $\rho$ meson with different lattice setups (right panel), including
different actions (overlap (O), clover (C), domain wall (D) and HISQ (H), different spacial lattice spacings and different pion masses. The information
of the configurations used can be found in Refs.~\cite{Blum:2014tka,Lin:2008pr,Bazavov:2009bb}. 24J and 16J are anisotropic lattices with $a_s/a_t=3.7$, so
their highest effective masses are higher than those of 24I and 48I that have similar spacial lattice spacings. To increase visibility, some points in the right panel are shifted slightly
in the horizontal direction.\label{meff}}
\end{figure}

To explore the reason why there is no contribution for $\nu\gtrsim2$ GeV in Fig.~\ref{M2} and Fig.~\ref{M3}, 
we calculate the effective mass from 2-point functions which is a quick way to check the highest energy of intermediate states that
our Euclidean hadronic tensor contains. The results are plotted in the left panel of Fig.~\ref{meff}.
We see that for either $u$ or $d$ quark, the highest effective mass is around 1 GeV, which means 
that there is simply no information of higher excitations for this particular case. This should be due to lattice artifacts, since
the lattice we are using has finite volume (resulting in discrete momenta and discrete spectrum), finite lattice spacing (an UV cutoff) and unphysical pion 
mass (unphysical multi-particle states).
To sort out the most important factor,
we calculate the effective mass of the $\rho$ meson with different lattice setups (right panel of Fig.~\ref{meff}).
The reason we choose to check $\rho$ meson is because the hadronic tensor involves two vector currents inserted between the nucleon states
and the correlator of $\rho$ can be treated as two vector currents inserted between the vacuum.
Although the exact value of how high we can reach in the $\rho$ meson case may not have much to do with the hadronic tensor case,
the fact that how lattice artifacts affect the effective mass should be relevant.
The legend of the figure shows the features of the setups. Each label in the legend has four parts: valence quark type/sea quark type (``O'' denotes overlap, ``D'' for domain wall, 
``C'' for clover and ``H'' for HISQ), spacial size plus a suffix which serves as an identifier, spacial lattice spacing, and sea pion mass.
It is easy to see that, for 24I and 48I~\cite{Blum:2014tka}, although the pion masses and volumes are not the same, the highest effective masses are similar, around 3 GeV.
For 24J and 16J~\cite{Lin:2008pr}, the spacial lattice spacings are similar to the ones of 24I and 48I ($\sim$ 0.1 fm), but the highest effective mass can be higher than 5 GeV, which is because
these two lattices are anisotropic and their temporal lattice spacings are about 0.035 fm.
Then, for 48H~\cite{Bazavov:2009bb} and the two setups of 32If~\cite{Blum:2014tka}, despite their different fermion actions and volumes, their highest effective masses are all about 5.5 GeV and their lattice spacings are $\sim$0.06 fm.
For the lattice with lattice spacing $\sim$0.045 fm (64H~\cite{Bazavov:2009bb}), the highest effective mass is close to 8 GeV.
This test shows that the lattice spacing is the most important factor in order to have the information of higher excitations. 
In view of this comparison, the HISQ lattice with lattice spacing $\sim$0.045 fm can be a better choice to reach $\nu>2$ GeV.

In this paper, we formulate our approach of calculating the hadronic tensor on the lattice. We point out that this is the approach that includes the inclusive 
contribution of all the intermediate states which is crucial to providing information for the neutrino scattering experiments at low energies.
It is also promising to calculate the structure function in the DIS region which can be used in the global fittings of parton distribution functions.
However, solving the inverse problem is the most challenging part. We have implemented and tested three algorithms using mock data, showing that the BR method has the
best resolution in extracting peak structures while BG and ME are more stable for the flat spectral function.
Realistic lattice results are presented for both the elastic case and a case with large momentum transfer.
For the elastic case, the reconstructed Minkowski hadronic tensor from the ME method gives precisely the vector charge which shows the feasibility of this approach.
For the latter case, the RES and possibly SIS contributions around 1 GeV are observed but no information is obtained 
for higher excited states with $\nu>2$ GeV.

A check of the effective masses of $\rho$ meson with different lattice setups indicates that, in order to reach higher energy transfers,
using lattices with smaller lattice spacings is essential for the lattice calculation.
The HISQ lattice with lattice spacing $\sim$0.045 fm
should be suitable to study the neutrino nucleus scattering at DUNE where the beam energy is between $\sim1$ to $\sim7$ GeV.
In the future, working on lattices with lattice spacing of 0.3 fm or smaller would be desirable for studying the parton physics.

\begin{acknowledgments}
KFL thanks X. Feng, A. Kronfeld, J. C. Peng, J. Qiu and Y. 
Hatta for 
illuminating and useful discussions.
We also thank the RBC and UKQCD Collaborations for providing their DWF
gauge configurations. This work is supported in part by the U.S. DOE
Grant No. DE-SC0013065 and DOE Grant No. DE-AC05-06OR23177 which is within the framework of the TMD Topical Collaboration.
This research used resources of the Oak Ridge
Leadership Computing Facility at the Oak Ridge National Laboratory,
which is supported by the Office of Science of the U.S. Department
of Energy under Contract No. DE-AC05-00OR22725. This work used Stampede
time under the Extreme Science and Engineering Discovery Environment
(XSEDE), which is supported by National Science Foundation Grant No.
ACI-1053575. We also thank the National Energy Research Scientific
Computing Center (NERSC) for providing HPC resources that have contributed
to the research results reported within this paper. We acknowledge
the facilities of the USQCD Collaboration used for this research in
part, which are funded by the Office of Science of the U.S. Department
of Energy. Y.Y. is also supported by the CAS Pioneer Hundred Talents Program.
\end{acknowledgments}

\bibliographystyle{apsrev4-1}
\bibliography{library}

\end{document}